\documentclass[referee]{aa}
\usepackage{graphicx}

\begin{document}

\title{X-ray properties of the transient pulsar 3A~0535+262
in quiescence}  

\titlerunning{3A~0535+262}

\author{U. Mukherjee \inst{}
\and
B. Paul\inst{}
}

\institute{Department of Astronomy $\&$ Astrophysics,
              Tata Institute of Fundamental Research,
              Homi Bhabha Road, Colaba, Mumbai-400 005,
              India  \\
              \email{uddipan@tifr.res.in, bpaul@tifr.res.in}
             }

   \date{ }

\abstract{
We present the timing and spectral properties of 
the transient Be/X-ray binary pulsar 3A~0535+262 
during quiescence using three observations with 
the narrow field imaging instruments (NFI) of BeppoSAX.
Assuming a distance of 2 kpc for this system, the 2--10 keV X-ray
luminosities measured from the three observations are in the range of 
1.5--4.0 $\times$ 10$^{33}$ erg s$^{-1}$, indicating a very low rate
of accretion. We report the detection of pulsations at a very low luminosity of 
2 $\times$ 10$^{33}$ erg s$^{-1}$ during one of the three 
observations, though at this accretion rate the system is expected to be in the 
centrifugally inhibited regime. The X-ray spectra for the unpulsed 
observations are best modeled as power law type while a combined model of 
power law and black-body is required to fit the pulsed spectrum.}

\maketitle


\section{Introduction}

3A~0535+262 is a close binary system which contains a O9.7 IIIe Be star
(HDE 245770, Giangrande et al. 1980) and a slowly rotating, strongly magnetized
neutron star. The X-ray source was discovered by the Ariel V satellite
(Rosenberg et al. 1975) when it was in a `giant outburst' mode with
X-ray intensity of the order of $\sim$ 2 Crab in the 3--7 keV range
and showed pulsations at 104 s. The system has been observed to undergo
several transient X-ray outbursts classified as ``giant'' 
(2--10 keV flux $>$ 1 crab) and ``normal'' (2--10 keV flux $< 1$ crab) 
outbursts. After 1975
two more giant outbursts were observed, one in 1980 detected with the
Hakucho satellite (Nagase et al. 1982) and another one in 1994 by the
BATSE detectors onboard CGRO (Finger et al. 1996). A large orbital period
of 111.0$\pm$0.4 days was discovered from the long term light curves
obtained with the Vela 5B satellite (Priedhorsky and Terrell 1983).
Interestingly, the system shows significant periodic optical variability
at a different period of $\sim$ 103 days (Larionov, Lyuty \& Zaitseva
2001). During the 1994 outburst, quasi-periodic oscillations (QPO) were
observed with frequency correlated to the X-ray flux and spin-up torque
(Finger et al. 1996, Ghosh 1996). The orbital parameters were also
determined from observations made with BATSE during this outburst.
The magnetic field strength of the neutron star in 3A~0535+262 is 
$\sim$ 1 $\times$ 10$^{13}$ Gauss as measured from the cyclotron resonance
scattering feature with the OSSE experiment onboard CGRO during the 1994
outburst (Grove et al. 1995).  

With EXOSAT observations made in 1985-1986 in between the outbursts,
a moderate and pulsating X-ray flux was detected from 3A~0535+262 at
various orbital phases (Motch et al. 1991). The X-ray flux in the
1--20 keV energy band was measured to be in the range of 1.5--3.0
$\times$ 10$^{-10}$ erg cm$^2$ s$^{-1}$, which corresponds to a luminosity
of 0.7--1.4 $\times$10$^{35}$ erg s$^{-1}$ for a distance of 2 kpc
(Steele et al. 1998). The detection of X-ray pulsations in two of
the EXOSAT observations showed that the mass accretion rate was large 
enough not to cause centrifugal inhibition 
and that material from the outflow of the Be star companion accretes
through a disk onto the neutron star.
In more recent observations of 3A~0535+262 made
with the Proportional Counter Array of the 
Rossi X-ray Timing Explorer (RXTE-PCA), 
the source was detected at a much lower luminosity 
of 2.0--4.5 $\times$ 10$^{33}$ erg s$^{-1}$ at which   
the onset of centrifugal inhibition of accretion or propeller effect 
is expected. The X-ray luminosity that was
still detected in the propeller regime was ascribed to material leaking
through the magnetosphere or thermal emission from the heated core of the
neutron star (Negueruela et al. 2000). However, as the RXTE-PCA detectors
are non-imaging instruments and therefore have large internal and X-ray sky
background, the flux and spectral measurements at low intensity are
not very accurate.

In this paper, we present X-ray timing and spectral properties of 3A~0535+262
in a quiescent state from three observations made with the Beppo-SAX narrow
field imaging spectrometer instruments. Detection of pulsations and measurement
of the X-ray spectrum in quiescent state may resolve the outstanding issue
regarding the different accretion regimes in which the system would reside.
In the subsequent sections we describe the observations (\S2), results from
timing and spectral analysis (\S3) and a discussion on the propeller effect 
in this pulsar and the X-ray spectral properties in the 
pulsating and non-pulsating states (\S4) based on the results obtained here.

\section{Observations}

Three observations of 3A~0535+262 were made with the Beppo-SAX observatory
on 2000 September 4 (5:14 UT), October 5 (00:42 UT) and 2001 March 5 (22:52 UT)
(observations A, B and C hereafter). The
scientific payload of Beppo-SAX consists of four Narrow Field Instruments
(NFI) comprising one unit of Low Energy Concentrator Spectrometer
(LECS; 0.1--10 keV, 22 cm$^{2}$ @ 0.28 keV) and three Medium Energy
Concentrator Spectrometer (MECS; 1.3--10 keV, 150 cm$^{2}$ @ 6 keV).
The energy resolution ($\%$ FWHM) of the LECS and the MECS detectors is
8 $\times$ (E/6 keV)$^{-0.5}$.  BeppoSAX also consists of a  High
Pressure Gas Scintillator Proportional Counter 
along-with a Phoswich Detection System.
A detailed description of Beppo-SAX
can be found in Boella et al. (1997).
Taking the time of periastron passage and the 
orbital period from Finger et. al. (1996), the orbital phases of the 
three observations A, B and C are 0.77, 0.05 and 0.42 respectively.
The respective observations are shown in Figure 1 along-with a sketch 
of the orbit of 3A~0535+262/HDE 245770 binary system.
The useful on-source times for the three observations were 32 ks, 39 ks and
51 ks each for the MECS and 19 ks, 30 ks and 43 ks for the LECS detectors.
As the source 3A~0535+262 was very
faint during these observations, we have analyzed data only from 
the narrow field imaging instruments. Data from only two of the MECS detectors
were available and the event files of the two detectors were added before
further analysis.

\section{Data Analysis}      

\subsection{Timing Analysis}

We extracted the MECS light curves with a time resolution of 1.0 s and
then corrected the arrival times to the times at the solar system 
barycentre. The average count rate of the light curves of A $\&$ B were 
$\sim$ 0.04 count s$^{-1}$ while that of C was 0.09 count s$^{-1}$.
All the three light curves showed intensity variations by more than
a factor of two at a few hours to day timescale. The light curve of C had
a stretch of 240 ks spread over several satellite orbits, which is useful for
accurate measurement of pulse period. We searched for pulsations in the
light curves of all the three observations
using the pulse folding and $\chi^{2}$ maximization technique in the 
expected pulse period range of 102--104 s with a
period resolution of 0.0035 s. Only the observation C
showed coherent pulsations with a period of 103.41 s and
the result of the period search is shown in  
Figure 2. Another period of 101.57 s also showed up when the 
period search was performed over a wider range between 101--105 s.
But by looking at the long term pulse period history of this pulsar (Figure 3),
it may be inferred that a period of 101.57 s is quite unlikely and
arose here due to the window function of the light curve.
 
To determine the lowest intensity in observation C 
at which pulsations can be
detected from 3A~0535+26, we subsequently divided the total MECS
observation duration into twenty four roughly equal segments in time.
Following this we extracted five separate light-curves each containing
data only from those segments with average count rate less than
0.15 count s$^{-1}$, 0.125 count s$^{-1}$, 0.1 count s$^{-1}$,
0.08 count s$^{-1}$, and 0.06 count s$^{-1}$ respectively. From
independent pulsation searches on these light curves with the same 
parameter space as mentioned above, we detected pulsations down to an
average count rate of less than 0.08 count s$^{-1}$.  
The inset in Figure 2 shows the period search result of the light-curve created
from the segments with average count rate less than 0.08 count s$^{-1}$. 
The light curve with segment-wise average count rate of less than 0.06
count s$^{-1}$ has a very small time duration and pulsations could
not be detected independently from it. However, if this
light curve is folded with the same pulse period as determined from
the entire observation, the pulse profile clearly shows intensity
variations correlated with the pulse profiles from the higher intensity
segments (Figure 4). The pulse profiles are background 
subtracted and the pulse fraction is [(Maximum - Minimum)/Maximum] $\sim$ 50$\%$. 
Thus it is fair to conclude 
that pulsations exist in 3A~0535+262 during observation C down to
at least a count rate of 0.06 count s$^{-1}$. Measurement of 
the X-ray luminosity of 3A~0535+262 corresponding to this count rate 
is presented in the next section.

\subsection{Spectral Analysis}

The source and background spectra were extracted from the MECS and
LECS detectors using circular regions of radius 4$\arcmin$ and 8$\arcmin$
respectively. The energy range chosen for MECS
was 1.8--10.5 keV while that for LECS was 0.3--4.5 keV, in which the
respective instruments have large effective area and the
spectral responses are well understood. For all the three observations,
the MECS and LECS phase averaged spectra were fitted simultaneously
with the relative normalization of the two instruments allowed
to vary. All the spectra were suitably binned to allow use of $\chi^{2}$
statistics. 

The continuum spectra of accreting X-ray pulsars are most often
described by a power law with a high energy cut-off, with the cut-off energy
usually above 10 keV. Therefore, while fitting a pulsar spectra in medium
energy band that is not very wide, only a power law or a black-body model is
invoked. Orlandini and Dal Fiume (2001) noted that the continuum of accreting
X-ray pulsars as observed by BeppoSAX can be well described in terms of a
black-body component, a power law and a high energy cutoff
($E_c$ $>$ 10 keV). The main physical process responsible for the
continuum emission in accreting X-ray pulsars is Compton up-scattering
of soft seed photons.
Depending on the value of the comptonization parameter(y),
the emergent spectrum of the photons can be a 
black-body type (y $\ll$ 1) or power law type with a high energy 
cut-off (if y $\gg$ 1 and the inverse compton scattering is unsaturated).
Motivated by these factors, we fitted the spectra in XSPEC v11.2.0 (Shafer,
Haberl and Arnaud 1989) with power law and black-body models  
along-with an absorption by material in the line of sight. 
The black-body model does not fit any one of the three 
spectra well and reduced $\chi^{2}$ of more than 2.0 were 
obtained. The results from the power law 
spectral fit are given in Table 1 along with 90\% confidence
limits for the spectral parameters. We found that the spectra from the
first two observations (A \& B) fit very well with   
the model while the spectrum from observation C during which the X-ray 
luminosity was higher, does not fit. 
Hence we tried to put in another additive component. 
A power law and a black-body component
along with absorption column density was found to fit the spectrum 
of observation C well and the spectral parameters are given in Table 2.
The power law photon index (for a single power law component for
observations A \& B and a double component model for observation C)
during these three observations is between 1.5--2.0, whereas the temperature
of the black-body component is 1.33 keV. Assuming a 2 kpc distance for the 
object, the radius of the emission region for this black-body model is estimated to
be $\sim$ 0.1 km. The absorption column density, for all the three observations
is about 6.5 $\times$ 10$^{21}$ atoms cm$^{-2}$, marginally above
the galactic column density towards 3A~0535+262 which is 5.9 $\times$ 10$^{21}$ 
atoms cm$^{-2}$. The 2--10 keV luminosities (in unit of 10$^{33}$ erg s$^{-1}$)  
for observations A, B \& C are 1.54, 1.62 and 4.07 respectively for power law
type spectra and 3.93 for the two component spectrum in observation C.
The spectrum from observation C is shown in Figure 5, along with the
model components and the residual to the best fit model.

The LECS and MECS spectra were extracted for the time intervals 
in observation C during which the average count rate was less than 
0.06 count s$^{-1}$ (see the previous section). These spectra were 
fitted simultaneously with
the power law plus black-body model with the absorption column 
density, power law photon index, black-body temperature, and the relative
LECS/MECS normalization being kept fixed to the values determined from the
entire observation. Thus the lowest luminosity at which pulsations
could be detected from 3A~0535+262 came out to be 
2.0 $\times$ 10$^{33}$ erg s$^{-1}$ calculated for a distance of 2 kpc.

\section{Discussion}

Transient X-ray pulsars in general are prone to wide variations in mass
accretion rates and are suitable systems to test the various accretion 
regimes onto high magnetic field neutron stars. They generally have
luminosities of the order of 10$^{32}$--10$^{33}$ erg s$^{-1}$ during 
quiescence. Detection of pulsations in these sources with such a low 
luminosity seems to be an interesting phenomenon regarding the classification
of the accretion regimes in which these systems may exist.

The luminosities obtained by us, 1.5--4.0 $\times$ 10$^{33}$ erg s$^{-1}$,  
during the three quiescence observations with BeppoSAX are similar to that
of Negueruela et al. (2000) measured with the RXTE-PCA in quiescence.
But the EXOSAT observations of this source during previous low states
(Motch et al. 1991) report a luminosity 
two orders of magnitude higher than what is observed in quiescence.
During observation C, we show that pulsations 
exist for a luminosity as low as 2.0 $\times$ 10$^{33}$ erg s$^{-1}$,
similar to the lowest luminosity obtained for this source in quiescence 
by Negueruela et al. (2000). However, as the RXTE-PCA detectors are
non-imaging instruments,
the diffuse X-ray emission from the galactic ridge can cause 
hindrance to an accurate measurement of the very low X-ray flux 
and the spectral properties of 3A~0535+262 in quiescence.

Assuming an efficiency $\eta$ = 1 in the conversion of gravitational energy into 
X-ray luminosity, the luminosity obtained by us in the whole of observation C 
( \L = 3.95 $\times$ 10$^{33}$ erg s$^{-1}$ ) 
gives an accretion rate \.{M} = 2.05 $\times$ 10$^{13}$ g s$^{-1}$.
For the known surface magnetic field strength and the spin 
period of this pulsar, we obtain the magnetospheric radius as 
\begin{equation}
r_{\mathrm m} = 9.35 \times 10^{9} ~~cm
\end{equation}
and the corotation
radius is calculated to be 
\begin{equation}
r_{\mathrm c} = 3.0 \times 10^{9} ~~~~cm
\end{equation} 
Therefore as r$_{\mathrm m}$ $>$ r$_{\mathrm c}$, 
following Stella et al. (1986), the pulsar in observation C 
as well as in A $\&$ B can be expected to be in the centrifugally 
inhibited regime. In this regime, accretion onto the neutron star 
surface is inhibited by the centrifugal action of the rotating magnetosphere 
(Illarionov and Sunyaev 1975). 
As observations A $\&$ B do not show pulsations, we infer that
at such a low level of luminosity they are indeed in the
centrifugally inhibited regime. During observation C, which 
has the highest luminosity among the three,
though r$_{\mathrm m}$ $>$ r$_{\mathrm c}$    
would mean the existence of a centrifugal barrier, pulsations have been
detected unambiguously. The detection of pulsations in C probably means that a 
fraction of the disk material is somehow going 
onto the surface of the neutron star along the magnetic field lines.
The three BeppoSAX NFI observations combined together show that at a
luminosity between 1.5--4.0 $\times$ 10$^{33}$ erg s$^{-1}$,
3A~0535+262 makes a transition between centrifugal
inhibition and direct accretion onto the neutron star surface.
We would like to note here that there are some uncertainties regarding
the calculation of r$_{\mathrm m}$.
If $\eta$ is not equal to 1 or if the distance to the source is very 
different from what has been assumed; the value
of r$_{\mathrm m}$ would change. A very low $\eta$ 
would correspond to higher accretion rate which 
in turn compresses the magnetosphere to give 
a smaller value of r$_{\mathrm m}$. But even with 
$\eta$ $\sim$ 0.1 the value of r$_{\mathrm m}$ 
does not come out to be less than r$_{\mathrm c}$. 
Similarly, a larger distance would imply a larger X-ray
luminosity and higher accretion rate.
The distance of 2 kpc to 3A~0535+262 has been
determined using spectral type and reddening of its companion star
(Steele et al. 1998) and is unlikely to be wrong by a large factor.
Another uncertainty in the calculation of r$_{\mathrm m}$ is
due to the assumption of aligned rotator.
Other X-ray transients like 4U 0115+63 (Campana et al. 2001
with BeppoSAX) and GRO J1744-28 (Wijnands et al. 2002 with Chandra) 
in quiescence with luminosities of $\sim$ 10$^{33}$ erg s$^{-1}$
have also been thought to be in the centrifugally inhibited regime. 

We found that the phase averaged spectrum for observation C 
fitted well when a 
combined model of power law and black-body along with absorption was used. 
Else in the spectra of the other two observations A \& B,
the power law index ($\Gamma$) was between 1.5--2.0 and the equivalent 
hydrogen column density obtained was of the order of 10$^{21}$ atoms cm$^{-2}$. 
The spectral parameters of Motch et al. (1991) are in conformity with ours
as they obtained photon index $\sim$ 1.66 while 
the column density was $\sim$ 10$^{21}$ atoms cm$^{-2}$.
The low column density measured for this source matches with
the amount of material expected in the line of sight from reddening.
The visual extinction A$_{v}$ $\sim$
2.7 mag (Giangrande et al. 1980) gives a column density of $\sim$
(4.8--6.0) $\times$ 10$^{21}$ atoms cm$^{-2}$ (Predehl and Schmitt 1995,
Gorenstein 1975).
On the other hand, Negueruela et al. (2000)
obtained values of the column density which are an order of magnitude 
higher than that obtained here with the BeppoSAX spectrometers. 
Since the energy range for RXTE-PCA was taken to be 3--20 keV,
the measurements of column density by Negueruela et al. (2000)
cannot be compared with the values obtained here with combined BeppoSAX
MECS-LECS which goes down to 0.3 keV. As the low energy part of the 
spectrum is measured quite well with BeppoSAX,
it can also be used to have some gross understanding about 
the wind density and the mass-loss rate of the companion 
star. After taking into account the galactic column density towards 
this source (5.9 $\times$ 10$^{21}$ atoms cm$^{-2}$) it is evident that 
the column density arising due to local material is very low; of the 
order of 10$^{20}$ atoms cm$^{-2}$. A very low column density 
would imply a very low mass-loss rate for the companion
star which can cause the low mass accretion rate as observed here.
To have an order of magnitude estimation of the parameters,
we have calculated the column density profile 
with orbital phase for the pulsar by considering a 
spherically symmetric stellar wind emanating from the companion 
and assuming a terminal wind velocity of $\sim$ 1000 km s$^{-1}$ 
and the inclination to the orbit was taken to be 25$^\circ$
(Okazaki and Negueruela 2001).
Assuming spherical accretion, a set of three curves were obtained, 
each for the observations A, B $\&$ C with the mass-loss rates being 
calculated from their respective luminosities. The effective column
density values (Observed column density -- Galactic column density)
as obtained from the power law fits were superimposed on this set of curves
as shown in Figure 6. The effective values of the column density are of the 
same order of magnitude as of the model. Thus it can be said that 
a low luminosity and low value of the measured local absorption column
density are consistent with a low mass loss rate from the companion star.

It is to be noted that the spectra for the pulsating and the 
non-pulsating states are different. From the three BeppoSAX observations,
it is clear that a power law spectral model fits the quiescence spectra
much better than a black-body model, although an additional black-body
component helps in improving the fit significantly for observation C. 
From RXTE-PCA spectrum (Negueruela et al. 2000) it was not possible to
distinguish between power law and black-body models.
The radii of the emitting region for the black-body component
in the third spectrum came out to be almost two orders of magnitude 
smaller compared to the canonical 10 km neutron star radius.  
It is plausible that at very low accretion rates as is the case here, 
the plasma not having enough ram pressure to compress the magnetosphere 
moves along the outermost field lines and gets deposited on the very
inner parts of the polar caps and hence occupy a rather small 
and compact region. In this scenario, the black-body component in 
the spectra of observation C with very low emitting region may be possible.
In this regard, we mention the RXTE observations of the persistent low
luminosity X-ray pulsar 4U 0352+309 (Coburn et al. 2001).
The area of the black-body component ($\sim$ 10$^8$ cm$^2$; same as obtained in obs. C) seen in
its spectra in low state was consistent with being emitted from the
magnetic polar cap of the pulsar.

This black-body component in observation C is quite different from 
the soft excess usually observed in some binary X-ray pulsars
(Yokogawa et al. 2000, Paul et al. 2002). They are in general modeled as a
black-body or thermal bremsstrahlung where the black-body temperature is 
$\sim$ 0.1 keV and the area of the emitting region $\sim$ 10$^{15-16}$ cm$^{2}$.
Recently, Hickox, Narayan and Kallman (2004) have explored 
the physical origin of this so called ``soft excess'' and found that for 
sources with luminosity L$_{x}$ $\leq$ 10$^{36}$ erg s$^{-1}$,
it may be due to emission by photoionized or collisionally
heated diffuse gas, or thermal emission from the surface 
of the neutron star. In the highly luminous (L$_{x}$ $\geq$ 10$^{38}$ erg s$^{-1}$)
sources, the soft component may be due to reprocessing of hard X-rays by the
optically thick, accreting material, most likely near the inner edge of the
accretion disk. At low luminosities, like in the present case, the inner
disk temperature can not be high enough for it to be an X-ray emitter.
Moreover, the size of the black-body emission region in 3A~0535+262 is very small compared to that of the
inner accretion disk, and the flux in the black-body component is a substantial
fraction of that in the power law component ($\sim$50\% in the present case). These facts 
suggest that the black-body component is not reprocessed by hard X-ray emission
as is the case in many luminous High Mass X-ray Binary (HMXB) pulsars. Rather, this soft black-body component
resembles that of 4U 0352+309 (Coburn et al. 2001).
In another HMXB, 4U 1700--37, a previously known soft X-ray excess 
(Haberl et al. 1994) has now been found to consist of a blend of low energy
emission lines (Boroson et al. 2003). However, in the cases of 3A~0535+262
(present work) and 4U 0352+309 (Coburn et al. 2001), the temperature of the soft
component is an order of magnitude higher and is unlikely to be a blend of narrow emission
lines.

\section{Conclusions}

We have investigated the timing and spectral properties of the X-ray
pulsar 3A~0535+262 in quiescent state using the narrow field instruments
of the BeppoSAX observatory. We have detected pulsations in observation C
in parts of which the 2--10 keV X-ray luminosity was as low as
2 $\times$ 10$^{33}$ erg s$^{-1}$. 
For all the three observations of 3A~0535+262 the accretion is expected to be 
centrifugally inhibited though the detection of pulsations 
at these low flux levels in one of 
them indicates that some matter may have leaked through onto the neutron
star surface. When the emission is non-pulsating, the X-ray spectrum is
power law type. The origin of an additional black-body component in the
pulsed observation is not very clear though it seems to be essential
in fitting the spectra.

\section{Acknowledgments} 
We thank the referee L. Kaper for many valuable suggestions, and also the
BeppoSAX team for providing the data in the public archive.
UM thanks K. Sengupta and S. Naik for their help in preparing the
manuscript and data analysis. He also acknowledges the Kanwal Rekhi
Scholarship of T.I.F.R. Endowment Fund for partial financial support.

\begin{figure}
   \centering
   \includegraphics[angle=-90,scale=0.6]{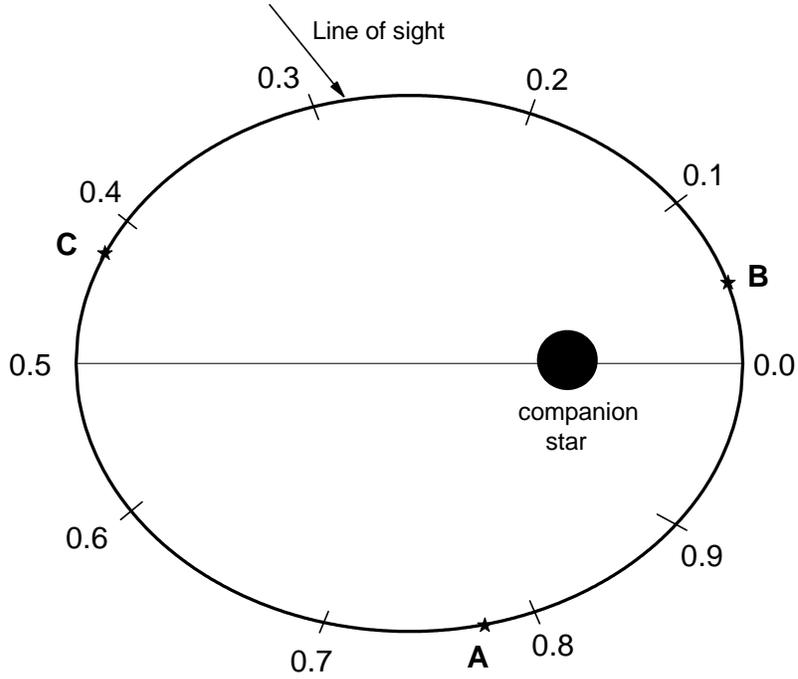}
   \caption{Orbital geometry of 3A~0535+262/HDE 245770 binary system. The orbital phases are marked on the 
circumference of the ellipse. The observations A, B and C are denoted with asterisks and the projection of the 
line of sight on the orbital plane is also shown. The orbital parameters are taken from Finger et al. (1996). The 
radius of the companion star is 15 R${_\odot}$ and the inclination of the orbit is $\sim$ 25$^\circ$
(Okazaki and Negueruela 2001) 
              }
\end{figure}

\begin{figure}
   \centering
   \includegraphics[angle=-90,width=12cm]{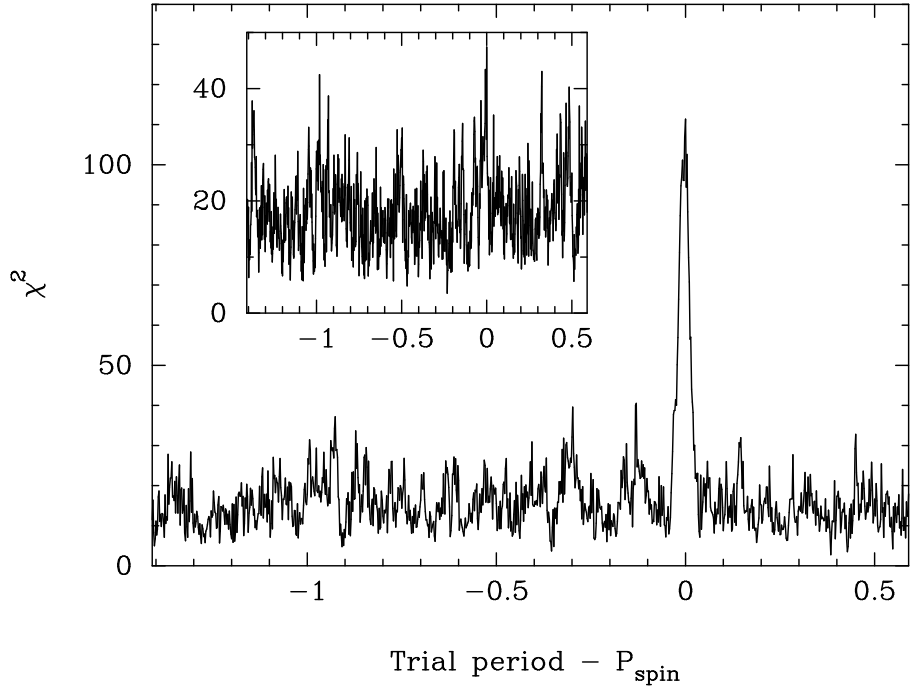}
   \caption{The results of the period search for the third observation  
with the inset showing the period search result for the portion of the lightcurve with average count
rate less than 0.08 count s$^{-1}$. 
              }
            \end{figure}

\begin{figure}
   \centering
   \includegraphics[angle=-90,width=10cm]{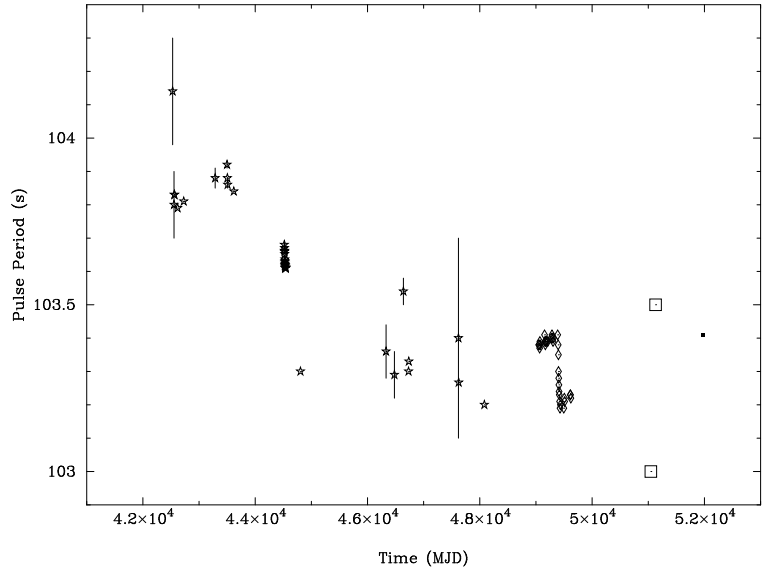}
   \caption{ The long term pulse period history of 3A~0535+262 taken from Coe et al. (1990) and other references 
cited in the text. The diamonds represent the BATSE observations by Finger et. al. (1996), the open squares show 
the two RXTE-PCA observations by Negueruela et al. (2000) and the filled square denotes the observation C 
with Beppo-SAX (MJD $\sim$ 51973).
              }
\end{figure}

\begin{figure}
   \centering
   \includegraphics[angle=-90,width=10cm]{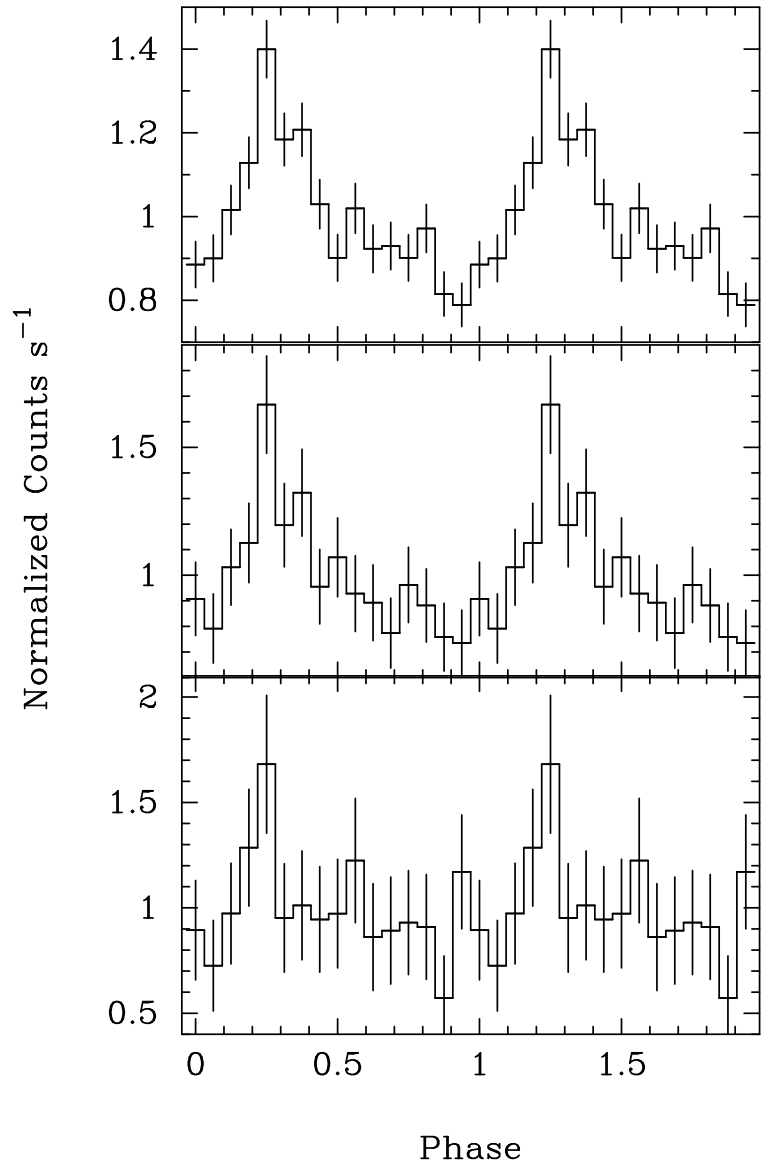}
   \caption{The top panel shows the pulse profile for the entire duration of observation C. The middle panel shows
the pulse profile for the part of the lightcurve of observation C in which the count rate is less than 
0.08 count s$^{-1}$ and the bottom panel shows the pulse profile of the part of the lightcurve of C in which the count 
rate is less than 0.06 count s$^{-1}$. 
              }
\end{figure}

\begin{figure}
   \centering
   \includegraphics[angle=-90,width=14cm]{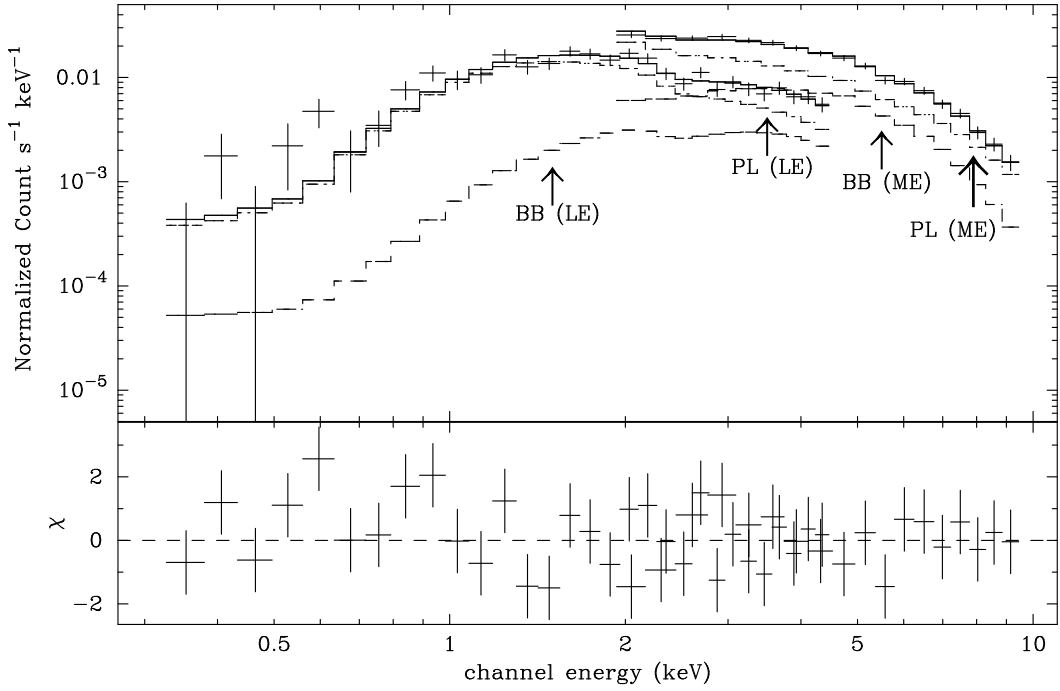}
   \caption{The fitted spectrum of observation C along with the residuals in the lower panel. The individual 
components of the spectral model are also shown (ME : MECS, LE : LECS, PL : Power Law and BB : Black Body)
              }
\end{figure}

\begin{figure}
   \centering
   \includegraphics[angle=-90,width=10cm]{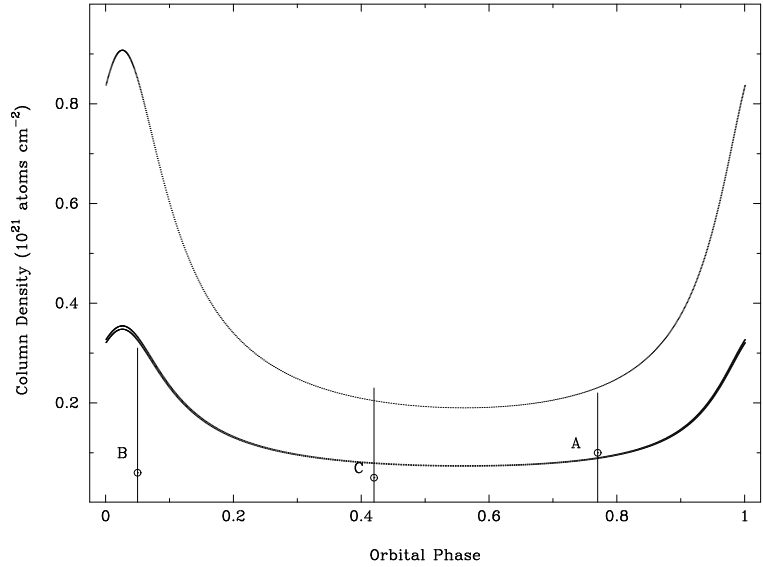}
   \caption{Model variation of column density with orbital phase calculated for three different mass loss-rates
corresponding to the three observations. The observed column densities after subtracting 
the galactic column density are superposed along with their error-bars.
              }
\end{figure}

\begin{table}[b]
\caption{Power Law Spectral parameters. The luminosities are in the 2--10 keV range for a distance of 2 kpc 
         }
~\\
\begin{tabular}{cccccccc}
\hline
Obs. &N$_H$(10$^{22}$cm$^{-2}$)  &Index($\Gamma$) &L (10$^{33}$ erg s$^{-1}$) &Red.$\chi^{2}$ (d.o.f.) \\
\hline
\hline
\\
A &$0.65^{+0.25}_{-0.12}$  &$1.93^{+0.17}_{-0.16}$ &1.54 &0.93 (44) \\
\\
\\
B &$0.64^{+0.18}_{-0.23}$ &$1.81^{+0.13}_{-0.13}$ &1.62 &1.07 (44) \\
\\
\\
C &$0.69^{+0.12}_{-0.12}$  &$1.69^{+0.07}_{-0.08}$ &4.07 &1.71 (44) \\
\end{tabular}
\end{table}

\begin{table}[b]
\caption{Power Law + Black Body Spectral parameters for Obs. C (The quoted luminosities are in the 2--10 keV range and 
assuming a distance of 2 kpc; 
L$_{BB}$ and L$_{PL}$ are the luminoisities corresponding to the Black body and Power law components.)}
~\\
\begin{tabular}{cccccccc}
\hline
N$_H$(10$^{22}$cm$^{-2}$)  &Index($\Gamma$) &kT(keV) &R(km) &L$_{BB}$ (10$^{33}$ erg s$^{-1}$) &L$_{PL}$ (10$^{33}$ erg s$^{-1}$) &Red.$\chi^{2}$ (d.o.f.) \\
\hline
\hline
\\
$0.6^{+0.6}_{-0.6}$  &$1.84^{+0.34}_{-0.18}$ &$1.33^{+0.17}_{-0.2}$ &0.08 &1.36 &2.57 &1.05 (42) \\
\end{tabular}
\end{table}
\end{document}